\documentclass[12pt,preprint]{aastex}

\shorttitle{Continuum modeling of the $\delta$~Sco disk}
\shortauthors{Carciofi et al.}

\begin{document}

\title{Properties of the $\delta$~Scorpii Circumstellar Disk from Continuum Modeling}

\author{A.~C.~Carciofi\altaffilmark{1}, A.~S.~Miroshnichenko\altaffilmark{2,3},
A.~V.~Kusakin\altaffilmark{4,5}, J.~E.~Bjorkman\altaffilmark{3},
K.~S.~Bjorkman\altaffilmark{3}, F.~Marang\altaffilmark{6},
K.~S.~Kuratov\altaffilmark{5}, P.~Garc\'\i a-Lario\altaffilmark{7},
J.~V.~Perea Calder\'on\altaffilmark{8},
J.~Fabregat\altaffilmark{9} and
A.~M.~Magalh\~aes\altaffilmark{1}
}

\altaffiltext{1}{Instituto de Astronomia, Geof\'isica e Ci\^encias
Atmosf\'ericas, Universidade de S\~ao Paulo, Rua do Mat\~ao 1226,
Cidade Universit\'aria, S\~ao Paulo, SP, 05508--900, Brazil}
\altaffiltext{2}{Department of Physics and Astronomy, University of
North Carolina at Greensboro, P.O. Box 26170, Greensboro, NC
27402--6170, USA} \altaffiltext{3}{Ritter Observatory, Department of
Physics and Astronomy, University of Toledo, Toledo, OH 43606--3390,
USA} \altaffiltext{4}{Sternberg Astronomical Institute,
Universitetskij pr. 13, Moscow, Russia} \altaffiltext{5}{Fesenkov
Astrophysical Institute, Kamenskoe plato, Almaty 480068, Kazakhstan}
\altaffiltext{6}{South African Astronomical Observatory, PO Box 9,
Observatory 7935, South Africa} \altaffiltext{7}{European Space
Astronomy Centre, Research and Scientific Support Department of ESA,
Villafranca del Castillo, Apartado de Correos 50727, E--28080
Madrid, Spain } \altaffiltext{8}{European Space Astronomy Centre,
Villafranca del Castillo, Apartado de Correos 50727, E--28080
Madrid, Spain} \altaffiltext{9}{Observatorio Astron\'omico,
Universidad de Valencia, 46100 Burjassot, Spain}

\email{carciofi@usp.br}

\begin{abstract}
We present optical $WBVR$ and infrared $JHKL$ photometric
observations of the Be binary system $\delta$~Sco, obtained in
2000--2005, mid-infrared (10 and $18\;\mu$m) photometry and optical
($\lambda\lambda$ 3200--10500 \AA) spectropolarimetry obtained in
2001. Our optical photometry confirms the results of much more
frequent visual monitoring of $\delta$~Sco. 
In 2005, we detected a significant decrease in the object's brightness, both in optical and
near-infrared brightness, which is associated with a continuous rise in the hydrogen line strenghts. We discuss possible causes for this phenomenon, which is difficult to explain in view of current models of Be star disks.
The 2001 spectral energy distribution and polarization
are succesfully modeled with a three-dimensional non-LTE Monte Carlo
code which produces a self-consistent determination of the hydrogen
level populations, electron temperature, and gas density for hot
star disks. Our disk model is hydrostatically supported in the
vertical direction and radially controlled by viscosity. Such a disk
model has, essentially, only two free parameters, viz., the
equatorial mass loss rate and the disk outer radius. We find that
the primary companion is surrounded by a small (7 $R_\star$),
geometrically-thin disk, which is highly non-isothermal and fully
ionized. Our model requires an average equatorial mass loss rate of
$1.5\times 10^{-9} M_{\sun}$\, yr$^{-1}$.

\end{abstract}


\section{Introduction}\label{intro}

Line emission in most stellar spectra arises from ionized gas beyond
the photospheric level. In particular, emission-line spectra of
B-type stars (e.g., Be, B[e], Herbig Ae/Be) are due to extended
circumstellar (CS) envelopes. Additionally, the ionized CS gas
produces continuum radiation due to free-free and free-bound
transitions \citep{geh74}. The envelope contribution to the object's
brightness depends on the distribution of the density, temperature,
and ionization degree throughout the envelope.

Classical Be stars are a large class of objects in which CS
contribution to the stellar continuum can be significant. These
variable non-supergiant stars were found to show transitions between
active (with line emission) and passive (or diskless, without line
emission) phases \citep{por03}. 
In their active phase, Be stars exhibit moderately strong emission-line spectra (equivalent width of the H$\alpha$ line can be up to $\sim$70Ê \AA) and display large IR excesses, which can be up to $\sim$40 times larger than the photospheric flux of the central star at 12$\mu$m \citep{iras88}.
It has also been shown that the H$\alpha$ line
strength correlates with the IR excess \citep{cc88}.

Given the high level of the IR CS continuum, a question of the
importance of the CS contribution to the optical continuum arises.
It is definitely lower than in the IR, since photospheric lines are
seen in the optical range and the observed optical color-indices are
not far from intrinsic ones. However, the effect of the envelope
parameters on the CS continuum has not been studied in any
significant detail. If the optical excess radiation is large, it may
affect at least the determination of the Be star luminosities and
the amount of their CS gas as well as parameters derived from the
line profile modeling. From a comparison with normal B-type dwarfs
and giants, \citet{zb91} showed that the optical CS excess radiation
can be as large as 1 mag and that it also correlates with the Balmer
line strength. This study used pre-Hipparcos distance determinations
and average $V$-band brightnesses that could have affected the
derived CS luminosity excesses, but it qualitatively demonstrated
the importance of the optical CS emission to the total flux.

Although optical photometric observations of Be stars are counted in
tens of thousands \citep[e.g.,][]{pav97,pb01}, simultaneous
photometric and spectral observations are still rare, and only a few
Be stars have been observed in both phases (active and diskless)
with both techniques. These data show, for example, that the optical
brightness of $\gamma$ Cassiopeae was about 0.7 mag fainter in a
diskless phase than in an active phase with the strongest H$\alpha$
emission \citep[see][]{hum98}. Observations of $\pi$ Aquarii, which
had the most active phase in late 1980's and has lost the line
emission completely in 1996, show that the $V$-band brightness
difference between the two phases was $\sim$0.5 mag \citep{b02}.
This study has shown that, in the diskless phase, $\pi$ Aqu has lost
its entire IR excess, indicating that the CS gas has vanished from
the system.

In the summer of 2000, one of the brightest B-type stars, $\delta$
Scorpii, exhibited first signs of line emission. Follow-up
photometric and spectroscopic monitoring \citep{g02,m03} showed that
the primary companion of this binary system had a steadily
developing CS disk. In particular, the system became $\sim$0.7 mag
brighter in the $V$-band. Its H$\alpha$ strength increased with
time, and an initially double-peaked line profile became
single-peaked in 2003 \citep{m03}.

Line profiles in Be stars are
still ambiguously interpreted by different models
\citep[e.g.,][]{hv00,c05}. Therefore, it is important to constrain CS parameters using photometric observations.
Unfortunately, only a few photoelectric photometric observations of
$\delta$~Sco in more than one band have been published so far
\citep{g02}.

This paper has two main goals. In \S~\ref{obs} we present
multiwavelength photoelectric photometry of $\delta$~Sco obtained in
2000--2005, and in \S~\ref{mod} we present a detailed modeling of the
CS disk of $\delta$~Sco using the 2001 spectral energy distribution
(SED) and polarization. In \S\S~\ref{discuss}--\ref{conclus} we
discuss our results and present our conclusions.

\section{Observations \label{obs}}

The photometric $WBVR$ observations of $\delta$~Sco were obtained in
2002--2005 at the Tien-Shan Observatory (TSAO, Kazakhstan) with a
50-cm telescope and a standard pulse-counting single-channel
photometer. The instrumental system, whose $BVR$ bands are very
close to the Johnson system and $W$-band is a shorter version of the
Johnson $U$-band, is described in \citet{kmm85}. The $W-B$
color-indices were transformed to the wider used $U-B$ using the
following relationship for early B-type dwarfs (B0--B3) that is
based on comparison of the Johnson photometry and data from the
original $WBVR$ catalog of \citet{kor91}: $U-B=(0.83\pm0.01)\,(W-B)
+ (0.04\pm0.01)$.

The Johnson $BVRIJHK$ observations were obtained in
2002 at the 1 m TSAO with the 2-channel photometer-polarimeter FP3U of the
Pulkovo Observatory \citep{b88}. The errors of individual
measurements do not exceed 0.03 mag in all the bands. HR\,5984
(B0.5 {\sc V}) and HR\,5993 (B1 {\sc V}) were used as comparison
stars at both TSAO telescopes. Stability of the instrumental system was controlled using
other standard stars observed during the night.

In 2000--2005 we obtained $JHK$ photometry of $\delta$~Sco at the
1.55-m Carlos S\'anchez Telescope (CST), operated by the Instituto
de Astrof\'\i sica de Canarias at the Spanish Observatorio del Teide
(Tenerife, Spain). We used a CVF infrared spectrophotometer equipped
with an InSb photovoltaic detector, operating at the temperature of
liquid nitrogen, with a photometric aperture of 15\arcsec\ and a
chopper throw of 30\arcsec\ in the E-W direction to subtract the
contribution from the background sky. The Teide photometric system,
as well as its relations with other standard photometric systems, is
described in \citet{ar87}. The $JHK$ photometry in 2005 was obtained
with the photometer FIN that replaced the CVS at the CST. FIN is
also equipped with a liquid nitrogen cooled InSb detector and works
in chopping mode. The aperture and chopper throw were 15$\arcsec$
and 25$\arcsec$, respectively. The measurement uncertainties does
not exceed 0.03 mag in all the bands.

Additional $JHKL$ photometric data were obtained at the
0.75--meter telescope of the South-African Astronomical Observatory
equipped with a single--element InSb photometer and calibrated with
the SAAO photometric system \citep{c90}. The optical ($V$-band) and
near-IR ($K$-band) light curves along with the amateur visual
brightness estimates are shown in Figure \ref{light_curve}.

The IR imaging was performed on October 22, 2001, at Mauna Kea with
the 3\,m NASA Infrared Tele\-scope Facility (IRTF) and the IR camera
MIRLIN. The camera has a 128$\times$128 pixel, high-flux Si:AS BIB
detector with a plate scale of 0$\farcs$475 at IRTF. Background
subtraction was carried out by chopping the secondary mirror at an
11 Hz rate with a 15$\arcsec$ throw in the north-south direction and
by nodding the telescope a similar distance in the east-west
direction. The observations were obtained in the $N$ (effective
wavelength $\lambda_{\rm eff}$=10.79 $\mu$m and passband $\Delta
\lambda$=5.66 $\mu$m) and $Q_s$ ($\lambda_{\rm eff}$=17.90 $\mu$m
and $\Delta \lambda$=2.00 $\mu$m) bands. Five cycles, each
consisting of 25 co-added 50 chop pairs, were carried out in the $N$
band. In the $Q_s$ band we also took 5 cycles, each containing 20
co-added 100 chop pairs. The closest standard star, HR 6147, was
observed 20 minutes after $\delta$~Sco, but at a higher
elevation that caused a relatively large uncertainties in the
measured brightness.

The imaging data were reduced using an IDL package developed at Jet
Propulsion Laboratory (Pasadena) for MIRLIN. The resultant image has
a 32$\times$32 pixel field. The total flux within the object's
images was measured using the PHOT task under the APPHOT package in
IRAF\thanks{IRAF is distributed by the National Optical Astronomy
Observatories, which are operated by the Association of Universities
for Research in Astronomy, Inc., under cooperative agreement with
the National Science Foundation}. The object's brightness was
calibrated using known brightness of the standard stars in the $N$
($\lambda_{\rm eff}$=10.8 $\mu$m, $\Delta \lambda$=5.6 $\mu$m) and
$Q_s$ ($\lambda_{\rm eff}$=17.9 $\mu$m, $\Delta \lambda$=1.7 $\mu$m)
bands and in the IRAS 12 and 25 $\mu$m bands. The IRTF photometric
system is defined in \citet{h98}. The brightness of $\delta$~Sco
turned out to be $0.42\pm$0.10 mag in $N$ band and $0.9\pm$0.2 mag
in $Q_s$ band.

Optical spectropolarimetry was obtained in April-July 2001 using the
half-wave polarimeter (HPOL) at the 0.9 m telescope of the
University of Wisconsin's Pine Bluff Observatory (PBO). For details
about the observing program, instrument, and data reduction, see
\citet{wnn96} and references therein. Observations were made
successively in two separate spectral ranges: blue (3190--6050 \AA)
and red (5960--10410 \AA).

\section{Modeling \label{mod}}

$\delta$~Sco has been a standard for spectral classification (spectral type B0
{\sc IV}) since long ago. Its photometric properties have also been
measured at different epochs and showed no obvious variations
\citep[see][for more information]{m01}. The object is a
non-eclipsing binary system with a $\sim$1.5 mag optically fainter
secondary companion with an orbital period of 10.6 years
\citep{bed93} and a highly eccentric orbit \citep[$e$=0.94,][]{m01}.
In order to determine the object's SED before the beginning of the
Be phase, we collected available optical and IR photometric data
from the literature. These include optical and near-IR photometry in
the region 0.3--4.8$\mu$m \citep{the86} and IRAS data at 12 and
25$\mu$m \citep{iras88}. 
Inspection of the resulting SED shows no deviation from the theoretical SED \citep{kur94} for the primary's parameters from Table 3. This indicates that the secondary's  T$_{\rm eff}$ is not significantly different from that of the primary. Therefore, we simply removed 15\% of the dereddened flux at all the wavelengths to construct the pre-active (diskless) phase SED of the primary,
shown as the asterisks in Figure~\ref{best_fit_F}.

We now have to choose an SED that is representative of the active phase.
The light curve of $\delta$~Sco in 2001--2004 is characterized by a somewhat well-defined plateau around $V\approx1.7\;\rm mag$, with several short fadings (timescales of weeks to
months, Figure~\ref{light_curve}). 
As will be discussed more fully in
\S~\ref{discuss}, these variations are probably a result of a
modification of the properties of CS disk, perhaps an increase of
the mass loss rate associated with a change in the geometry.
We assume that the fadings are transient perturbations of a more
static disk configuration which is associated with the $V=1.7\;\rm mag$ plateau, 
and is represented by a somewhat smooth
density distribution of the CS gas. We, therefore, construct the
active phase SED from the highest fluxes obtained by us.
The active-phase SED is shown as the filled circles in Figure~\ref{best_fit_F}.


\subsection{Disk Model}

Observational evidence supports the idea that the disks of classical
Be stars are Keplerian (rotationally supported) gaseous disks
\citep[see][for a recent review]{por03}. The essential physics that
determines the geometrical structure of Keplerian disks is
reasonably well understood. The disk radial structure is governed by
viscosity \citep*{lee91}, while the vertical structure is controlled
by gas pressure. 


Assuming a steady-state isothermal outflow, the disk density
structure is given, in cylindrical coordinates $(\varpi,z,\phi)$, by
\begin{equation}
\rho(\varpi,z) = \frac{\Sigma(\varpi)}{\sqrt{2\pi}H(\varpi)}
\exp\left(-\frac{z^2}{2H^2}\right).
\label{eq:density}
\end{equation}
\citep[see, for example, ][]{bjo97}. 
Using the viscosity prescription of \citet{sha73}, the disk surface density, $\Sigma$, is given by
\begin{equation}
    \Sigma(\varpi)=\frac{{\dot M} v_{\rm crit} R_{\star}^{1/2}}
           {3 \pi \alpha a^2 \varpi^{3/2}}
           \left(\sqrt{R_d/\varpi}-1\right)  ,
\label{eq:disk_Sigma}
\end{equation}
where $\alpha$ is the viscosity parameter.
The density scale is controlled primarily by the equatorial mass loss rate, $\dot{M}$, but also by the critical rotational velocity of the star, $v_{\rm crit}$, and the disk temperature via the sound speed, $a$. For isothermal disks with large $R_d$, the surface density is a power-law function of the radial distance, $\Sigma\propto \varpi^{-2}$.

The disk scale-height, $H$, is given by
\begin{equation}
                H(\varpi)=(a/v_\phi)\varpi  . \label{eq:scale_height}
\end{equation}
Since the disk is isothermal and the $\phi$ component of the velocity is Keplerian ($v_\phi=v_{\rm crit} \varpi^{-1/2}$), we find the familiar result that an isothermal disk flares as $H\propto\varpi^{1.5}$. It follows from Eqs.~(\ref{eq:density}) and (\ref{eq:scale_height}) that, for isothermal disks, the density falls very sharply with radius as $\rho \propto \varpi^{-3.5}$.
Another important property of Keplerian disks can be readily derived from Eq. (\ref{eq:scale_height}). Since $v_\phi$ is much larger than the sound speed (the first is of the order of several hundreds of km/s and the second a few tens of km/s), the disk scaleheight is small compared to the stellar radius  and the disk is geometrically thin.

To conclude this brief description of the properties of isothermal Keplerian viscous disks, we must describe their velocity structure. The $\phi$ component is Keplerian, as stated above, but those disks must have an outflow velocity, which is given by the mass conservation relation
\begin{equation}
            v_\varpi = \frac{\dot M}{2 \pi \varpi \Sigma}\;. 
            \label{eq:radvel}
\end{equation}

Recently, \citet[][hereafter CB]{car05a} presented a new
three-dimensional non-LTE Monte Carlo (MC) code for solving the
radiation transfer and radiative equilibrium problem for arbitrary
gas density and velocity distributions. The code was used to study
the thermal properties of the disks around classical Be stars.

CB found that the optically thick part of Be star disks are highly non-isothermal,
with kinetic temperature ranging from about 30\% of the stellar
effective temperature ($T_\mathrm{eff}$) in the midplane to
temperatures close to $T_\mathrm{eff}$ near the stellar surface.
CB also found that the disks are generally fully
ionized, but for later spectral types (B5 and later) the midplane
can become neutral, depending on the disk density scale.


From Eqs.~(\ref{eq:density}) to (\ref{eq:scale_height}), it is evident that the temperature plays an important role (via de sound speed) in the disk density structure; therefore, a self-consistent solution for the disk density, that takes into account all the non-LTE and three-dimensional radiative transfer effects on the disk temperature, is required to accurately predict the properties of the CS disk. 
This was done by 
\citet{car06}
who built upon the previous code and presented a self-consistent solution for the
disk density \citep[see also][]{bjo05}. 
Given a prescription for the viscosity
\citep[e.g.,][]{sha73} the disk model has only two free parameters,
the disk mass loss rate 
and outer radius.
All other quantities (surface density, vertical density and outflow
speed) are determined self-consistently from the solution of the
fluid equations. 
In this paper, we use an updated version of the CB code to model the active-phase SED and polarization of the primary companion of $\delta$ Sco.

 \subsection{Best-fitting Model}

The parameters of the primary companion
were taken from \citet{m01}
and are summarized in Table~\ref{t3}. We assume that the disk around
the primary companion is in the orbital plane of the binary system,
so the disk inclination angle is the same as the orbital inclination
angle $i$, and is a fixed parameter in our modeling within the
observationally derived error of $\pm\;5$\degr. Also, we assume a value of 0.1 for the viscosity parameter of \citet{sha73}.

\begin{figure}
\plotone{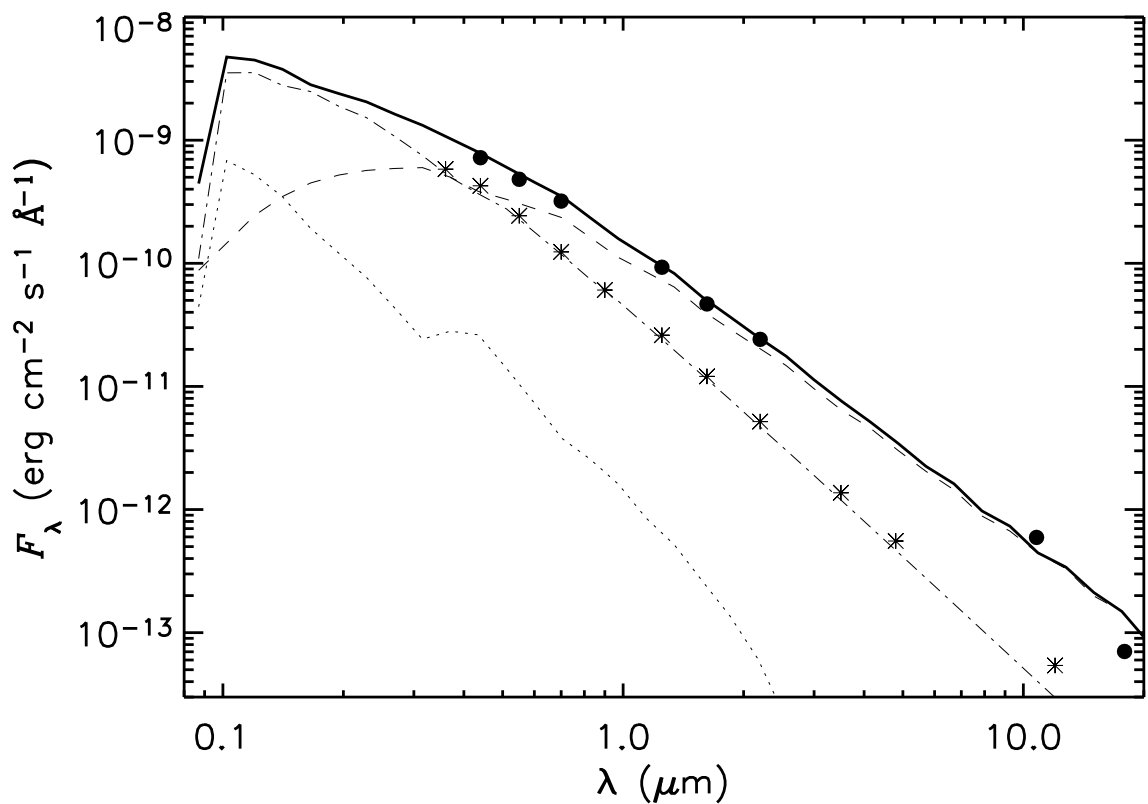} \caption{Best-fitting model. Our best-fitting SED
for the active phase (solid line) is shown along with the observed
SED (filled circles). The pre-active phase photometric data taken
from the literature also plotted (asterisks) for comparison. The
other lines correspond to the scattered, emitted and unprocessed
stellar fluxes (dotted, dashed, and dash-dot lines, respectively). }
 \label{best_fit_F}
\end{figure}

Figure~\ref{best_fit_F} shows our best-fitting SED  along with the
observed data.
We see that our model reproduces well the observed SED for all
wavelengths. Our best-fitting model consists of a very dense
(midplane density at the stellar surface, $\rho_0$, of $4.5\times
10^{-10}\, {\rm g\,cm}^{-3}$) but small ($R_d=7R_\star$) disk. In
\S~\ref{prop} we discuss in more detail the properties of our solution for the disk. The parameters of our best-fitting model are listed
in Table~\ref{t4}.

There is a number of interesting points to consider in
Figure~\ref{best_fit_F}. From the $J$-band longward ($\lambda >
1.25\;\mu\rm{m}$) the SED is completely dominated by the disk
emission and the stellar flux corresponds to only 1--20\% of the
total flux, depending on the wavelength. The stellar flux dominates
the spectrum only for $\lambda \lesssim 4000\rm\AA$, but the disk
emission still contributes at an important level down to $\lambda
\sim 2000\rm\AA$.

Another interesting feature of the SED is that, in addition to the usual
IR excess, there is a small UV excess just longward of the Lyman
jump. This UV excess is a result of electron scattering in the disk,
which redirects some of the stellar flux into the polar direction.

Finally, we call attention to the fact that the unprocessed stellar
radiation (\emph{dash-dot} line in Figure~\ref{best_fit_F}), which corresponds to the SED that the object would have
without the disk, matches very well the pre-active SED. This is an
important verification of our disk model, because it shows that it
predicts correctly the excess flux emitted  by the CS material.

\begin{figure}
\plotone{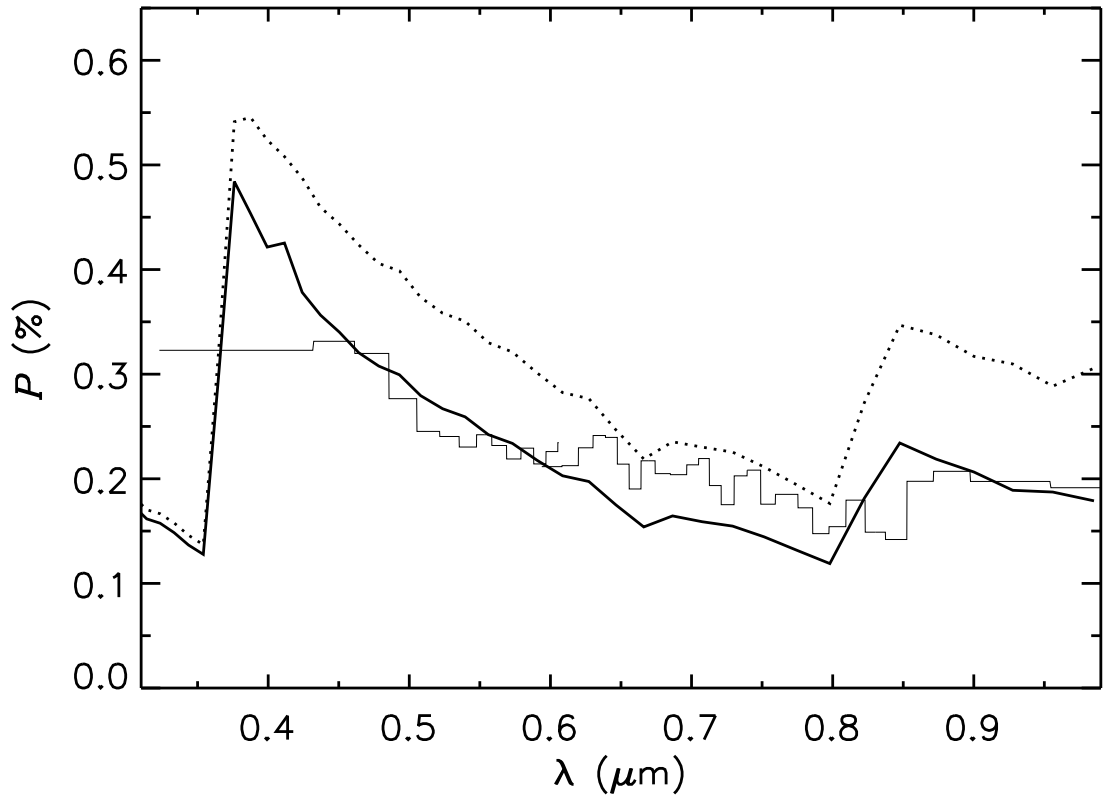} \caption{Best-fitting model. The best-fitting model
polarization (thick line) and the observed polarization (thin line)
are shown. The solid thick line correspond to the model polarization
with the depolarizing effect of the secondary included, and the
dotted thick line correspond to the uncorrected polarization.}
 \label{best_fit_P}
\end{figure}

The polarization results are shown in Figure~\ref{best_fit_P}.
Because of the large error of the observed polarization in some
wavelength regions, we bin the data so that the wavelength size of
the bin changes to keep a constant polarization error of 0.01\%.
Even using this procedure, we still observe a large scatter in the
data, which makes us believe that the actual errors in the bins may
be larger.

The observed polarization was corrected for the interstellar
contribution according to the measurement by \citet{hal58}. The
later ($B$-band polarization degree of 0.3\% and polarization angle
of 118$^{\circ}$) was most likely obtained in the early 1950's,
during the diskless (i.e., intrinsically unpolarized) phase.

Since we do not include the secondary in our modeling, the model
polarization (dotted line in Figure~\ref{best_fit_P}) must be
changed in order to account for the unpolarized flux of the
secondary. The depolarized curve is shown as the thick solid line of
Figure~\ref{best_fit_P}. 

We see that there is a general good
agreement between the model and the observations. For the blue part
of the Paschen continuum ($\lambda \lesssim 0.6 \mu \rm m$), both
the shape and level of the polarization are correct, but our model
predicts a somewhat smaller polarization for the red part.
Unfortunately, the observations were not able to resolve the Balmer
jump, but they do resolve the Paschen jump at $0.82\;\mu\rm m$ and
there is a rough agreement in the size of the jump. This is a
particularly important result since the Paschen jump in polarization
is controlled by the ratio of the $n=3$ and $n=4$ level populations,
and the agreement between the model and the observations indicates
that our non-LTE level populations are correct. Finally, we point
out the good agreement of the polarization in the Brackett continuum
($\lambda > 0.82\;\mu\rm m$).

\begin{figure}
\plotone{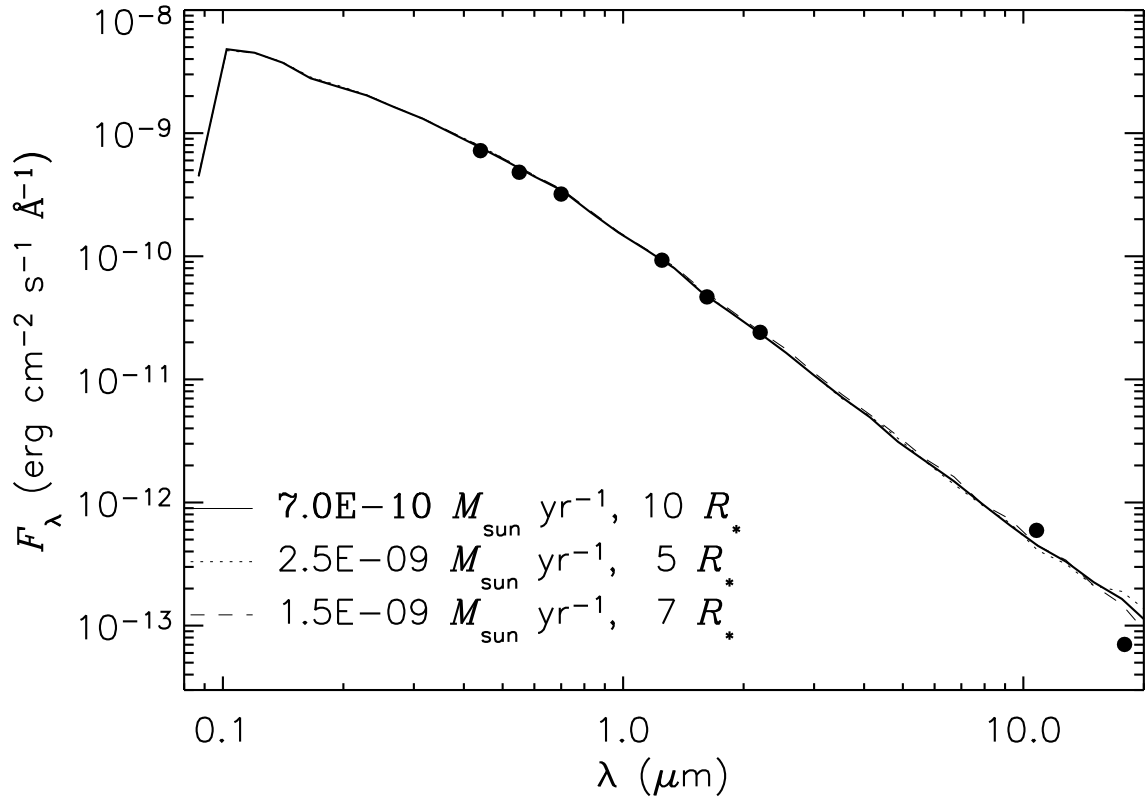} \caption{Comparison between the SED of different models, as indicated. The circles show the observed active-phase SED.}
 \label{unique_F}
\end{figure}

Let us now consider the uniqueness of our results for the disk
parameters. For the case of dusty circumstellar shells, it is a
well-known issue that fitting the SED alone can not well constrain
the dust properties and spatial distribution
\citep*[e.g.,][]{car04,mir99}. A similar issue is true here,
because, as illustrated in Figure~\ref{unique_F},  there are several
models that reproduce the observed SED equally well. For example, a
larger and less dense disk
($R_d=10\;R_\star$ and $\dot{M}=7\times10^{-10}\;M_\sun\,{\rm
yr}^{-1}$) is essentially equivalent to a smaller but denser disk
($R_d=5\;R_\star$ and $\dot{M}=2.5\times10^{-9}\;M_\sun\,{\rm
yr}^{-1}$).

\begin{figure}
\plotone{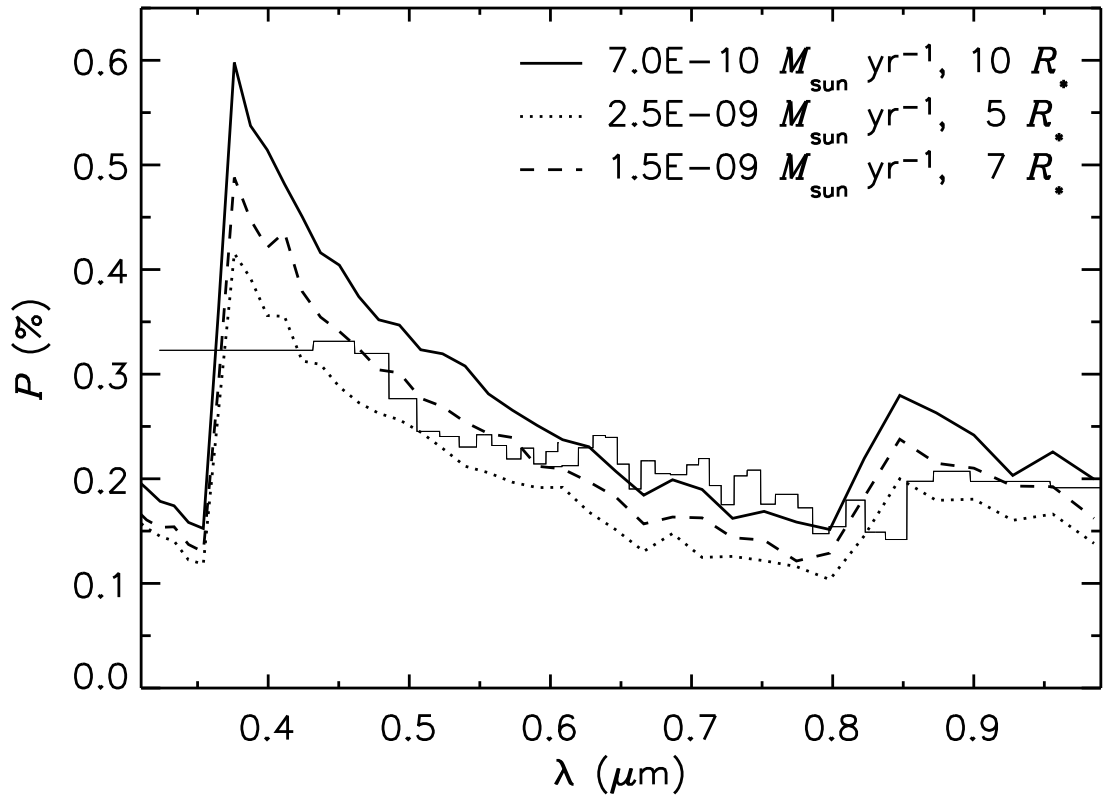} \caption{Comparison between the polarization of different models, as indicated. The thin line shows the observed polarization.}
 \label{unique_P}
\end{figure}

The linear polarization depends much more strongly on the model parameters ($\dot{M}$ and $R_d$) and, consequently, allows for breaking the degeneracy of the SED, as
illustrated in Figure~\ref{unique_P}. The differences between the
polarization levels of the three models shown are large enough to
allow one to choose the 7 $R_\star$ model  as to one that, for the
most part, best reproduces the data.




\subsection{Properties of the Solution} \label{prop}

\begin{figure}
\plotone{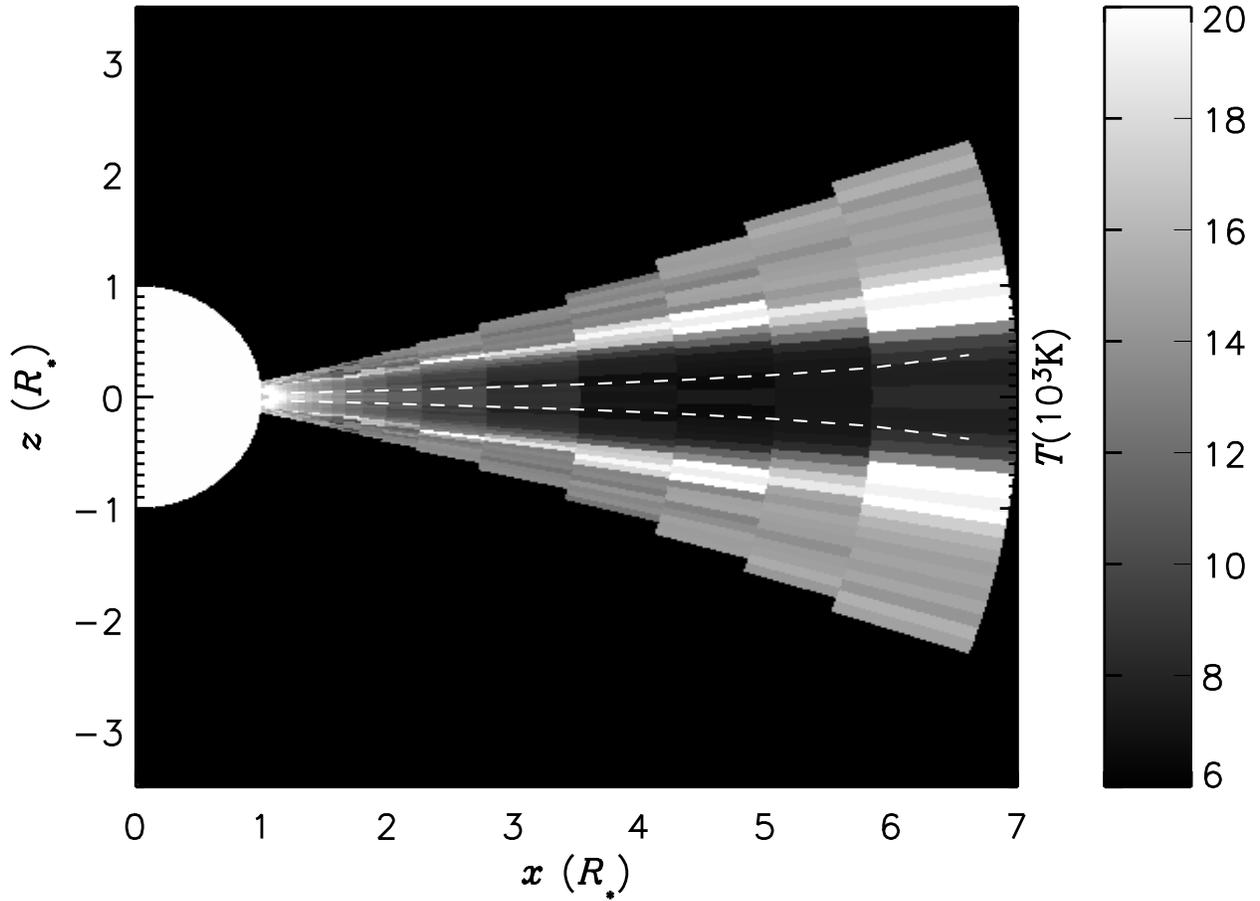} \caption{Temperature distribution of the
best-fitting model. The plot shows the temperature as a function of
$x$ and $z$. The dashed lines correspond to the curves
$z = \pm H(x)$ and show that the denser parts of the disk are geometrically very thin.}
 \label{temp}
\end{figure}

In this section we describe the properties of our solution for the
structure of the disk of $\delta$~Sco.
The temperature structure for our best fit model is
shown in Figure~\ref{temp}. CB showed that the disks of Be stars
are highly non-isothermal; indeed, in our results for $\delta$~Sco,
the temperature goes from a minimum of 7000 K at the midplane, around
$4R_\star$, to a maximum of about 30,000 K near the base of the disk.

As discussed in detail in CB, the
temperature behavior is a mixture between that of optically thick disks of
young stellar objects and optically thin winds of hot stars. For
the optically thin upper layers of the disk (defined as the regions where
the distance to the midplane is larger than one scaleheight), the
temperature is nearly isothermal, with an average of
about $0.6\;T_\mathrm{eff}$ or 16,000 K. At the midplane, however,
the temperature has a complicated structure. At the base of the disk,
the temperature is even larger than $T_\mathrm{eff}$ due to
back-warming of the star. The temperature drops very quickly with radius and its
profile is well-described by a flat blackbody reprocessing disk
\citep[][see also CB for more details]{ada87}.

\begin{figure}
\plotone{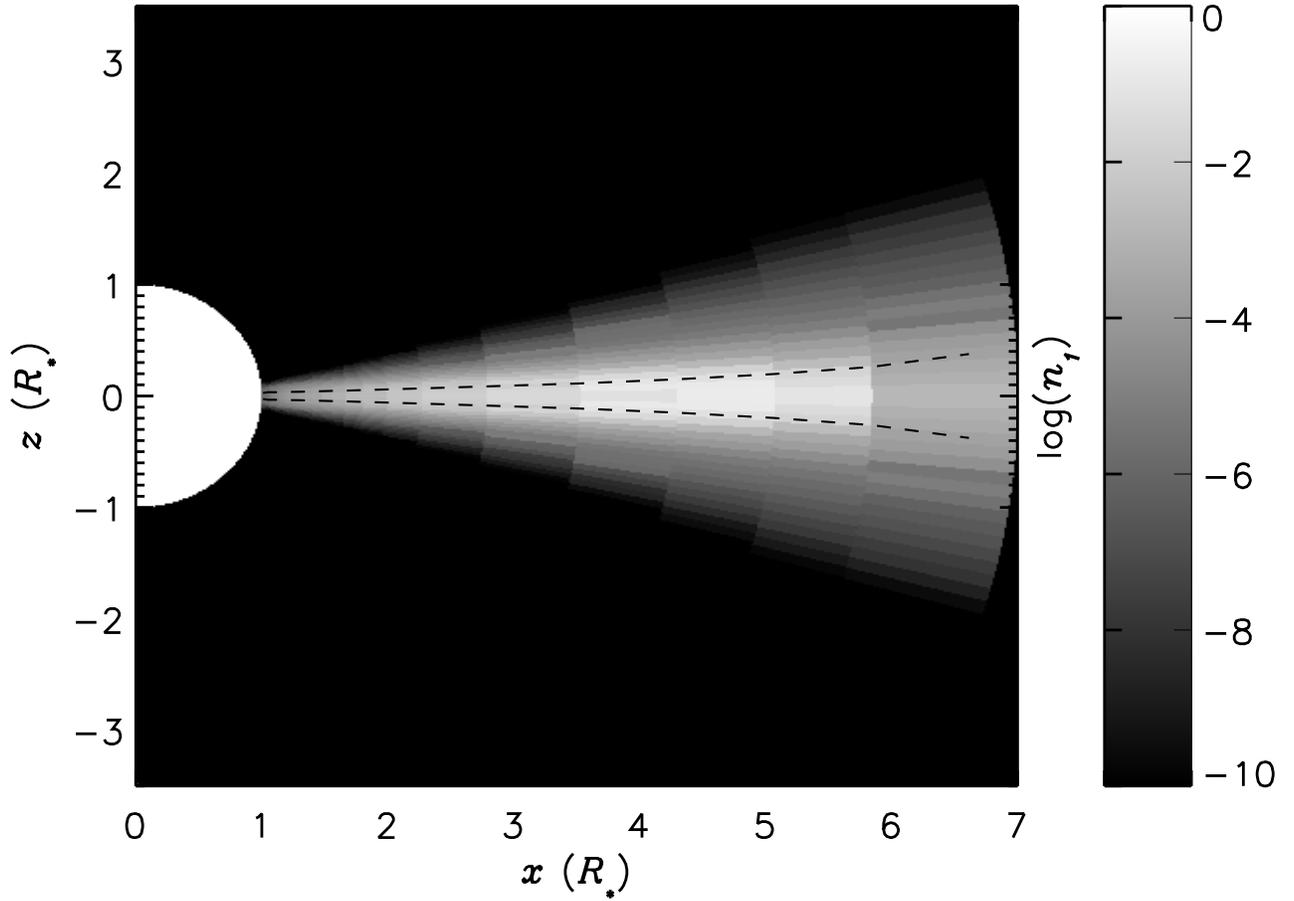} \caption{$n=1$ level populations for the
best-fitting model. The plot shows the logarithm of the fractional
level population as a function of $x$ and $z$. For a neutral gas,
$\log(n_1)=0$. The dashed lines correspond to the curves
$z = \pm H(x)$.}
 \label{level}
\end{figure}

In Figure~\ref{level} we show a map of the $n=1$ hydrogen level
population. We see that the disk is fully ionized, but there is a
minimum in the ionization fraction ($\approx 90\%$) that coincides
approximately with the minimum of the kinetic temperature. This
ionization minimum occurs because of the deficit of photoionizing
radiation due to the large optical depths of the midplane.

\begin{figure}
\plotone{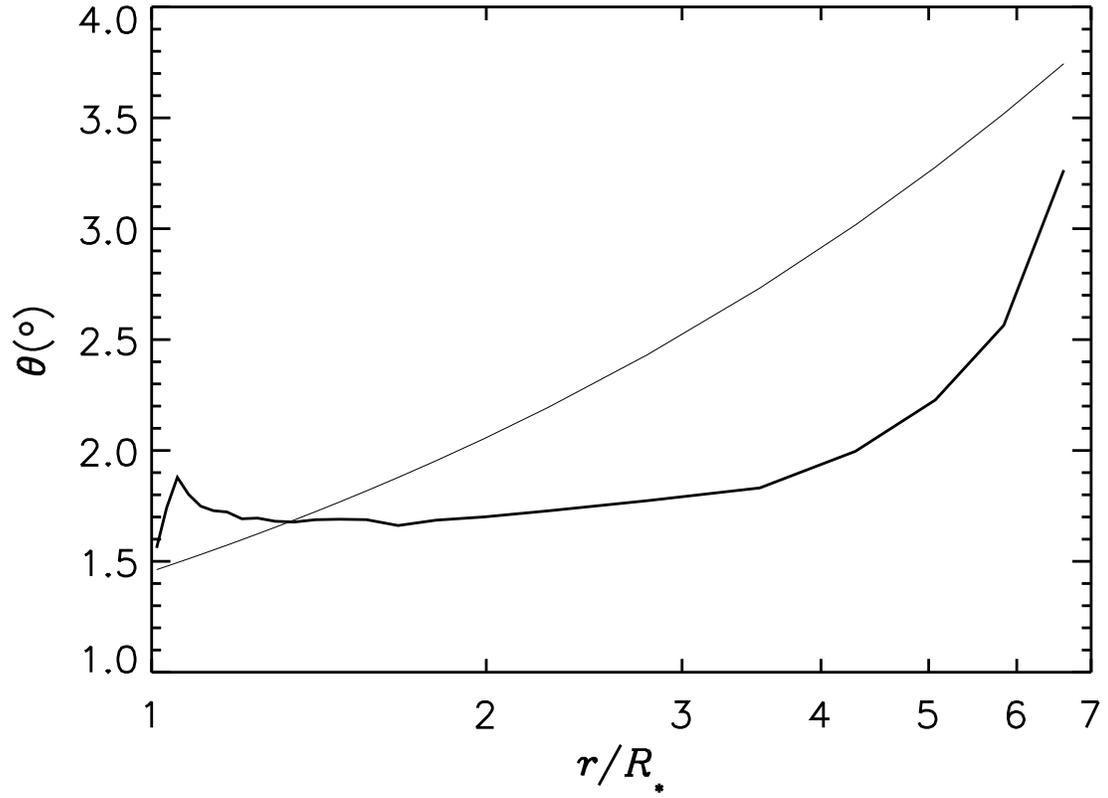} \caption{Opening angle of the best-fitting model.
Shown is the opening angle [thick line, eq.~(\ref{opangle})] along
with the opening angle of a corresponding isothermal model with
$T=16000\;\rm K$ (thin line).}
 \label{angle}
\end{figure}

Following \citet{car06}, we use our solution for
the disk temperature to self-consistently solve the fluid equations
and determine the disk density. 
Let us analyze 
how our solution
differs from the isothermal case of eqs.~(\ref{eq:density}) and
(\ref{eq:scale_height}). In Figure~\ref{angle} we show the disk opening
angle as a function of radial distance, defined as
\begin{equation}
\theta = \tan^{-1}\left[\frac{H(\varpi)}{\varpi}\right]. \label{opangle}
\end{equation}
For reference we also show the opening angle of a corresponding
isothermal disk with $\beta=1.5$. As expected, the opening
angles are very small and the entire disk is geometrically very
thin. As a result of the small inclination angle of the system and
the small vertical extent of the disk, the disk covers very little
of the star and most of the hemisphere facing us is visible. This is
one of the reasons why the polarization levels are small.
In Figure~\ref{angle} we see an interesting result:
the opening angle is approximately constant in the inner disk,
in striking contrast with isothermal disks that flare
significantly. This result is explained by the rapid fall of the
disk temperature with radial distance.

\begin{figure}
\plotone{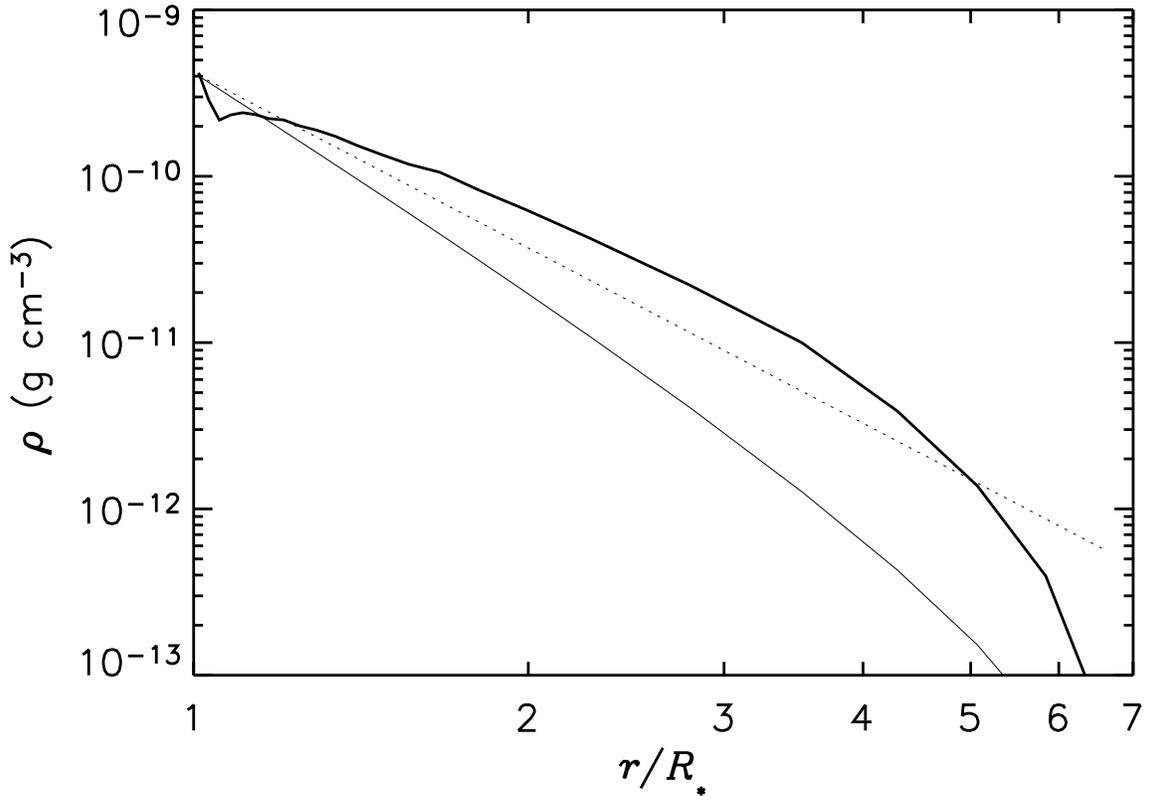} \caption{Density profile of the best-fitting model.
We compare the density calculated from our MC simulation (thick
line) to the $\rho \propto \varpi^{-3.5}$ curve expected for an
isothermal model (thin line).}
 \label{index}
\end{figure}

Let us, now, analyze the radial structure of the disk density, shown in
Figure~\ref{index}. We show, for comparison, the $\rho \propto
\varpi^{-3.5}$ curve corresponding to the isothermal solution for
the density. We see that the slope of the MC density profile for the
inner part of the disk departs significantly from the $-$3.5 value;
it is much less steep close to the star and is much steeper for $
\varpi/R_{\star}\gtrsim 3$. This is, again, a result of the
temperature structure in the midplane \citep[see][for detais]{car06}.

\section{Discussion} \label{discuss}

The disk model described above has only two free parameters: the
equatorial mass loss rate and the disk size. All the macroscopic
(density distribution, radial component of the gas velocity) and
microscopic (hydrogen level populations and gas kinetic temperature)
properties of the disk are self-consistently determined from those
two parameters. Since the observables critically depend on the CS
disk properties, both microscopic and macroscopic, it is of
significance that our two-parameter model can successfully reproduce
several different observations of $\delta$~Sco.

The polarization, for instance, is critically dependent on two
quantities: the geometry of the inner disk and the hydrogen level
populations. We have seen above that the geometry of the inner disk
is greatly altered by the temperature structure, with an interesting
result that the inner disk is essentially unflared. Another quantity
that is critically dependent on the details of the solution is the
slope of the IR SED, which depends on the radial density profile, as
demonstrated by \citet{w86}.

We believe that our model provides a good description of the
\emph{average properties} of the CS disk of $\delta$~Sco, which,
according to our assumptions, are associated with the highest
brightness levels of the system in 2001. As we discuss below, our
model cannot explain the large optical fadings of $\delta$~Sco
because a dynamical model, with a variable and possibly asymmetrical
mass loss, is probably required.

\begin{figure}
\plotone{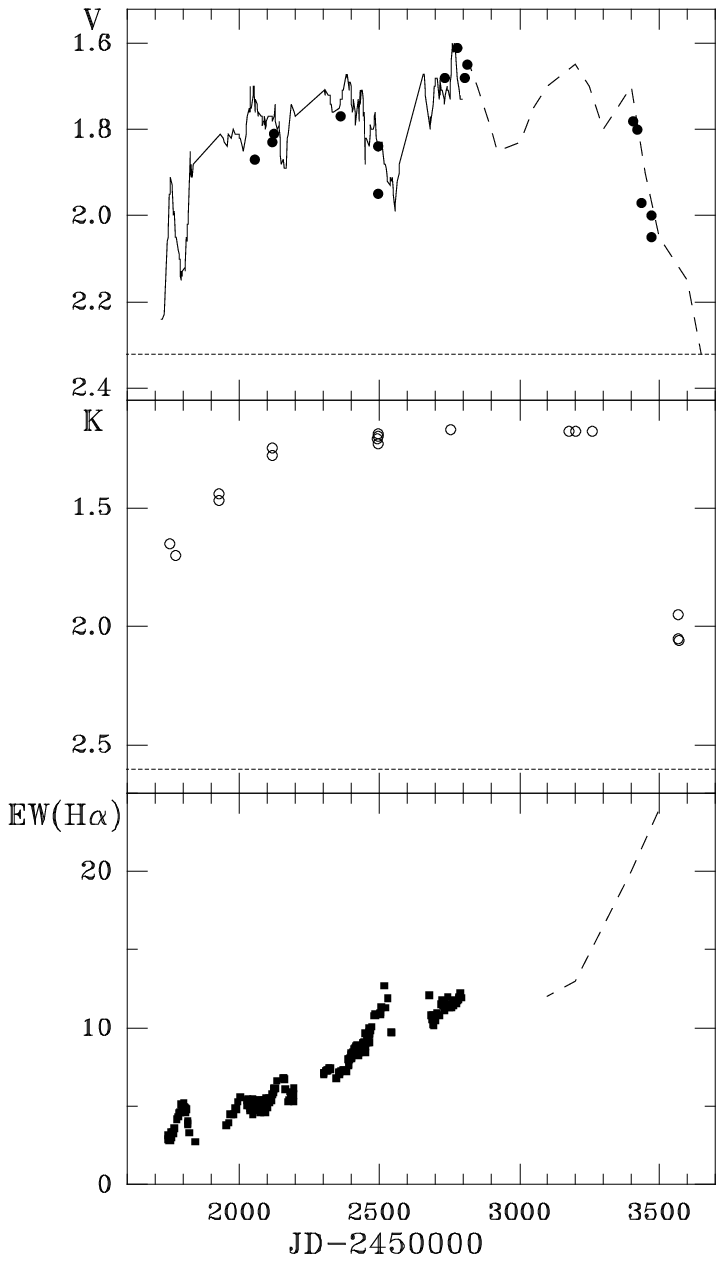} 
\caption{The optical (top panel) and near-IR (middle panel) light curves and the H$\alpha$ line equivalent width (EW, bottom panel) variations of $\delta$~Sco. Our photometric data are shown as circles (top panel), and the spectroscopic data from \citet{m03} are shown as filled squares (bottom panel). The solid line represents the visual light curve published by \citet{m03}, the long-dashed line in the top panel represents later data from 
http://ar.geocities.com/varsao/delta\_Sco.htm, 
the dashed line in the bottom panel represents low-resolution data from 
http://www.astrosurf.org/buil/becat/dsco/dsco\_evol.htm, 
and the short-dashed lines in each panel represent the diskless phase brightness level.}
 \label{light_curve}
\end{figure}

Our optical photometric data show that
visual estimates of the $\delta$~Sco brightness collected by
amateurs \citep[e.g.,][]{g02} are generally correct (Figure
\ref{light_curve}). This is important for future interpretation of
the entire active phase of $\delta$~Sco with dynamical models, which
can take advantage of the detailed optical light curve.

Our modeling shows that the CS disk of $\delta$~Sco, about 2 years
after the beginning of its formation, can be described as typical for
Be stars. 
The disk is geometrically thin in the vertical direction and
optically thick near the mid-plane.
At the same time, the optical light curve (Figure \ref{light_curve})
shows that a large flux excess 
($\Delta V \sim 0.7\;\rm mag$), due to continuum emission from the disk,
 was achieved very quickly, in about 2 years. This is not
typical of Be stars, since previous monitorings of their active
phase show that the optical brightness raises generally slower. It
took $\pi$ Aqr about 2 decades \citep{b02} and even longer to
$\gamma$ Cas \citep{telt93} to reach its highest optical brightness.
Other important difference between the light curve of $\delta$~Sco
and the historical light curves of other Be stars is that the latter
shows a good correlation between the emission-line strength (mostly
H$\alpha$) and the optical brightness, which is not observed for
$\delta$~Sco (see below).

Another feature of the light curve of $\delta$~Sco is the presence
of multiple optical fadings that last from weeks to months. A
striking feature is the pronounced fading in 2005, both in the optical
and near-IR wavelengths, that was accompanied by a continuing rise of the
line emission\footnote{The spectroscopic data have not been
analyzed yet. Preliminary results can be found at
\url{http://www.astrosurf.org/buil/becat/dsco/dsco\_evol.htm}.}
(see bottom panel of Figure~\ref{light_curve}).
This is a very unusual feature since other well-observed Be stars do not exhibit significant fadings,
unless it is a transition to a diskless phase (like that in $\pi$
Aqr).

\citet{m03} noted this anti-correlation between the visual and
H$\alpha$ fluxes in the earlier evolution of the $\delta$~Sco disk
and suggested that it might be explained by episodes of an increased
mass loss. Such matter ejections would produce an increase in the H$\alpha$
line emission strength which, in turn, would be followed by fadings
of the optical brightness due to an increase in the disk optical
depth.

This explanation for the optical fadings, however, raises more
questions than it answers. In a system such as $\delta$~Sco, part of
the optical flux comes from the star itself and part from the CS
disk, so there are basically two ways to reduce the optical
brightness: 1) to block part of the stellar disk; or, 2) to reduce,
via some mechanism, the disk emission. 

Since $\delta$~Sco has a low
inclination angle ($i \approx 35$\degr) and the CS disk is
(probably) geometrically very thin, this means that only about 10\%
 of the stellar flux can be blocked by the CS material, supposing a completely
 opaque disk in the vertical direction. It follows that, if process 1, above, is responsible for the large (0.3 mag or more)
fadings observed in $\delta$~Sco light curve, \emph{the
geometry of the system must be changed} to explain the large
attenuation of starlight. A possibility would be a radiatively
warped disk, as proposed by \citet{por98}. Such a warp would,
effectively, place material at higher latitudes and increase the
area of the stellar disk that could be blocked by the CS material.
Another possibility is to have material to be ejected from the star not at the
midplane, but at high latitudes. This ejection should, however, be
nearly continuous in the timescale of the optical fading (weeks or
months), since we expect the material to fall into the disk in an
orbital timescale, which is of the order of one to a few days.

If, on the contrary, process 2 above is responsible for the
fadings, the implications are somewhat similar. In our model,
material is injected into the disk at the equator and slowly
diffuses outward; therefore, an increase in the mass loss rate would result in
an increase of the optical recombination radiation. Thus,  a
modification of the geometry of the system is required in this case
also, if one wants to explain the optical fadings associated with
increased line fluxes observed for $\delta$ Sco.

We conclude, from this short phenomenological discussion, that our
static model of the $\delta$~Sco CS disk cannot explain the
optical fadings and their anticorrelation with the H$\alpha$ line
strengths. We suggest that a more complex model with a different CS
geometry is required to do that. It must be emphasized that
polarization is a very important constraint on the disk geometry,
since changes in the geometry would, most likely, result in
variations of the polarization angle. This quantity, therefore, is
of importance for constraining future dynamical models of
$\delta$~Sco.



Another result of our modeling is that we require a disk size of 7
$R_\star$. This small value makes sense in view of the recent
formation of the disk. \citet{m03} estimated the disk size based  on
the H$\alpha$ profile parameters, assuming both Keplerian rotation
($v_\phi \propto \varpi^{-1/2}$) and angular momentum
conservation ($v_\phi \propto \varpi^{-1}$), and found a disk
size of $8R_\star$ for the first case and $2.8R_\star$ for the
second. These spectroscopic estimates (averaged from the published
ones for the beginning of 2001 and beginning of 2003) are based on a
static approach to circumstellar gaseous rings by \citet{h72}. It is
interesting to note that the estimate assuming Keplerian rotation is
very similar to the value that comes from our modeling.


\section{Conclusions} \label{conclus}


We have obtained multicolor photometric observations (2000--2005)
along with the entire optical range spectropolarimetry (2001) of the
binary Be star $\delta$~Sco. Using these data, we constructed the
system's SED in the range 0.4--18 $\mu$m that roughly
corresponds to the maximum observed brightness.

Our photometric data confirm the visual light curve of $\delta$~Sco
that is being carefully observed by amateur astronomers
\citep[see][]{g02}. It also traced a deep fading of the object in
2005, almost all the way to its diskless phase brightness level. This
phenomenon is observed along with a continuous increase of the
object's emission-line spectrum that makes it highly unusual.
Similar anti-correlations of the brightness and emission-line
strength on a smaller time- and amplitude-scale have been observed
several times in 2000--2003 \citep{m03}. We suggest that this
phenomenon may be explained by a significant change both in the mass loss
rate and in the CS disk geometry.
Further frequent photometric, spectroscopic, and polarimetric
observations all the way to 2011, when the next periastron will
occur, are very important for understanding of the disk formation
and evolution in this highly eccentric Be binary.

We also modeled the observed SED, simultaneously with the linear
polarization, using a new three-dimensional non-LTE Monte Carlo
radiation transfer code. This code assumes that the disk is
hydrostatically supported in the vertical direction and that its
radial structure is governed by viscosity. The disk temperature and
density structure are solved self-consistently given only two input
parameters for the disk: the equatorial mass loss rate, which is
assumed constant, and the disk outer radius. The basic parameters of
the primary's CS disk are shown in Table \ref{t4}. We find that a
mass loss rate of $1.5\times 10^{-9} M_{\sun}$\, yr$^{-1}$ and a
disk size of $7R_\star$ best reproduces the available data.

Our solution for the disk temperature structure shows that, as have
already been demonstrated by CB, the disk is highly nonisothermal,
with temperatures ranging from 30\% of the stellar effective
temperature at the midplane, a few stellar radii away from the star,
to temperatures of the order of the stellar temperature at the base
of the disk. We also find that the disk is completely ionized.

The complex temperature structure of the disk has important
consequences on the disk density structure. We show that our
solution for the disk density departs significantly from the
isothermal solution. Most noticeably, we find that the inner disk is
essentially unflared, in contrast with the large flaring expected
for isothermal models. Also, we find that the radial density profile
departs significantly from the simple $n=-3.5$ power-law predicted
for isothermal models. Since the slope of the IR SED is mainly
controlled by the radial density structure, we suggest that careful
measurements of the IR SED, mainly in the near IR, may be used to
map the disk density and test the predictions of our model.

The next step in our modeling, which is already in progress, is to
include the synthesis of hydrogen line profiles. This new quantity,
associated with the SED and polarization, will help further
constrain the disk parameters.

\acknowledgements{A.~C.~C. and A.~M.~M. acknowledge support from the
S\~ao Paulo State Funding Agency FAPESP (grants 01/12589-1 and 04/07707-3).
A.~C.~C. and J.~E.~B. acknowledge support from NSF grants AST-9819928
and AST-0307686. 
A.~S.~M. and K.~S.~B. acknowledge support from NASA
grant NAG5--8054 and thank the IRTF staff for their assistance
during the observations. 
P~.G.~L. and J.~V.~P.~C.~ acknowledge support from grant
AYA-2003-09499 from the Spanish Ministerio de Ciencia y Tecnolog\'\i a.
A.~M.~M. acknowledges support from CNPq.
We thank the PBO observing team, and especially Marilyn Meade, Brian Babler, and Ken Nordsieck, for their invaluable assistance with obtaining, calibrating, and reducing the HPOL spectropolarimetric data.
This research has made use of the SIMBAD database operated at CDS,
Strasbourg, France.}

\clearpage

\clearpage

\begin{deluxetable}{cccrcc}
\tablecaption{Optical photometry of $\delta$~Sco \label{t1}}
\tablewidth{0pt}
\tablehead{
\colhead{MJD} &  \colhead{$V$} & \colhead{$U-B$} & \colhead{$B-V$} & \colhead{$V-R$} &
\colhead{$V-I$}
}
\startdata
2056.27& 1.87 &$-$    &$-$0.06&0.12&$-$\\
2117.15& 1.83 &$-$    &   0.04&0.27&$-$\\
2123.13& 1.81 &$-$    &   0.06&0.23&$-$\\
2362.40& 1.77 &$-$0.98&   0.12&0.24&$-$\\
2496.13& 1.95 &$-$0.91&   0.11&0.24&0.31\\
2497.14& 1.84 &$-$1.02&   0.14&$-$&$-$\\
2733.40& 1.68 &$-$0.88&   0.04&0.20&$-$\\
2779.25& 1.61 &$-$0.85&   0.12&0.29&$-$\\
2805.21& 1.68 &$-$0.88&   0.04&0.24&$-$\\
2815.17& 1.65 &$-$0.81&   0.08&0.33&$-$\\
3407.50& 1.78 &$-$0.91&   0.01&0.19&$-$\\
3420.48& 1.80 &$-$0.82&   0.00&0.17&$-$\\
3436.44& 1.97 &$-$0.86&$-$0.02&0.17&$-$\\
3471.40& 2.00 &$-$0.82&$-$0.01&0.21&$-$\\
3472.36& 2.05 &$-$0.86&$-$0.02&0.16&$-$\\
\enddata
\end{deluxetable}

\begin{deluxetable}{cccccc}
\tablecaption{Near-IR photometry of $\delta$~Sco \label{t2} }
\tablewidth{0pt}
\tablehead{
\colhead{MJD} &  \colhead{$J$} & \colhead{$H$} & \colhead{$K$} & \colhead{$L$} & \colhead{Obs.}
}
\startdata
1751.38& 1.77 &1.73&1.65&$-$ &Tenerife\\
1772.38& 1.86 &1.78&1.70&$-$ &Tenerife\\
1926.80& 1.67 &1.58&1.44&$-$ &Tenerife\\
1927.80& 1.66 &1.59&1.47&$-$ &Tenerife\\
2117.41& 1.46 &1.39&1.25&$-$ &Tenerife\\
2118.38& 1.48 &1.41&1.28&$-$ &Tenerife\\
2492.11& 1.53 &1.40&$-$ &$-$ &TSAO\\
2493.11& 1.54 &1.40&1.21&$-$ &TSAO\\
2495.10& 1.53 &1.44&1.20&$-$ &TSAO\\
2496.13& 1.54 &$-$ &1.23&$-$ &TSAO\\
2497.14& 1.60 &$-$ &1.19&$-$ &TSAO\\
2753.55& 1.45 &1.33&1.17&0.90&SAAO\\
3178.47& 1.48 &1.35&1.18&0.91&SAAO\\
3200.38& 1.47 &1.36&1.18&0.91&SAAO\\
3262.23& 1.48 &1.35&1.18&0.87&SAAO\\
3566.41&  2.14 & 2.07 &   1.95 &$-$ &Tenerife\\
3567.40&  2.19 & 2.15 &   2.05 &$-$ &Tenerife\\
3571.42&  2.21 & 2.16 &   2.06 &$-$ &Tenerife\\
\enddata
\end{deluxetable}

\begin{deluxetable}{cccccc}
\tablecaption{Stellar Parameters for $\delta$~Sco \label{t3}}
\tablewidth{0pt}
\tablehead{
\colhead{$R_{\star}$} &
\colhead{$T_\mathrm{eff}$} &
\colhead{$M_{\star}$} &
\colhead{$V_\mathrm{crit}$} &
\colhead{$i$} &
\colhead{Distance} \\
\colhead{$(R_{\sun})$} & \colhead{(K)} & \colhead{$(M_{\sun})$} &
\colhead{(km\,s$^{-1}$)} & \colhead{(\degr)} & \colhead{(pc)} }
\startdata
7 & $27,000$ & 14 & 620 & $38\pm5$ & 123 \\
\enddata
\end{deluxetable}

\begin{deluxetable}{llll}
\tablecaption{Best-fit Disk Parameters \label{t4}}
\tablewidth{0pt}
\tablehead{
\colhead{$\dot{M}$} & \colhead{$\rho_0$} &  \colhead{$R_d$} & \colhead{$i$} \\
\colhead{$(M_{\sun}\,{\rm yr^{-1}})$} & \colhead{$(\mathrm{g}\,{\rm
cm}^{-3})$} & \colhead{$(R_{\star})$} & \colhead{(\degr)} }
\startdata $1.5 \times 10^{-9}$ &
$4.5 \times 10^{-10}$  & 7 & 35 \\
\enddata
\end{deluxetable}

\end{document}